\documentclass[conference,final]{IEEEtran}

\IEEEoverridecommandlockouts

\usepackage{cite}
\usepackage{amsmath,amssymb,amsfonts}
\usepackage{algorithmic}
\usepackage{tikz}
\usepackage{graphicx}
\usepackage{textcomp}
\usepackage{xcolor}
\usepackage{siunitx}
\def\BibTeX{{\rm B\kern-.05em{\sc i\kern-.025em b}\kern-.08em
    T\kern-.1667em\lower.7ex\hbox{E}\kern-.125emX}}
\usepackage{environ}
\usepackage{ifdraft}

\ifoptionfinal{
    \usepackage[deletedmarkup=xout,commentmarkup=uwave,final]{changes} 
}{

    \usepackage[deletedmarkup=xout,commentmarkup=uwave,authormarkup=none]{changes}  
    \definechangesauthor[name={Thalia}, color=blue]{TP}
    \definechangesauthor[name={Dimitris}, color=blue]{DF}
}

\NewEnviron{addedblock}{%
    \ifoptionfinal{%
        \BODY
    }{
        {\color{blue}\BODY}%
    }%
}

\NewEnviron{deletedblock}{%
    \ifoptionfinal{%
    }{
        {\color{red}\BODY}%
    }%
}

\begin{document}


\newsavebox{\fundinglogo}

\sbox{\fundinglogo}{%
    \begin{tikzpicture}[baseline, overlay]
        \node[anchor=center] at (6em,-6pt) {\includegraphics[width = 0.4\columnwidth]{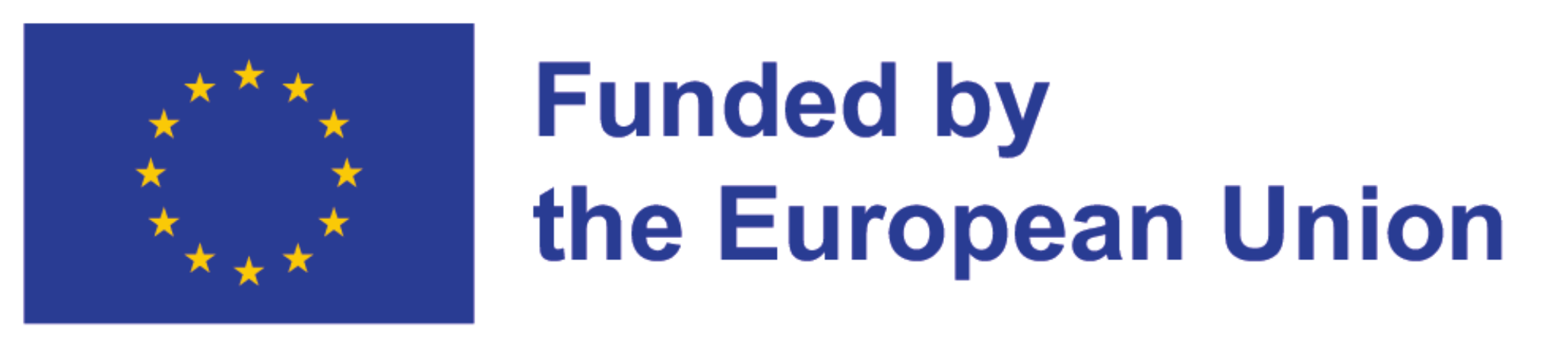}}; 
    \end{tikzpicture}%
}

\title{Reconfigurable Graphene-Metasurface Analysis via an Eigenmode-Free Method-of-Lines Formulation
\thanks{Copyright (C) 2026 IEEE. Personal use of this material is permitted.  Permission from IEEE must be obtained for all other uses, in any current or future media, including reprinting/republishing this material for advertising or promotional purposes, creating new collective works, for resale or redistribution to servers or lists, or reuse of any copyrighted component of this work in other works.\\ This project has received funding from the European
Union's Horizon 2020 research and innovation programme
under the Marie Sklodowska-Curie grant agreement
No.101146306. 
\protect\usebox{\fundinglogo}
}
}

\author{\IEEEauthorblockN{Maria-Thaleia Passia}
\IEEEauthorblockA{
{Electrical and Computer Engineering} \\
\textit{Aristotle University of Thessaloniki}\\
Thessaloniki, Greece\\
passiamg@ece.auth.gr}
\and
\IEEEauthorblockN{Dimitris Floros}
\IEEEauthorblockA{
{Nicholas School of the Environment} \\
\textit{Duke University}\\
NC, USA \\
dimitrios.floros@duke.edu
}
\and
\IEEEauthorblockN{Traianos V. Yioultsis}
\IEEEauthorblockA{
{Electrical and Computer Engineering} \\
\textit{Aristotle University of Thessaloniki}\\
Thessaloniki, Greece \\
traianos@ece.auth.gr}
}

\IEEEaftertitletext{\vspace{-2\baselineskip}}

\maketitle

\begin{abstract}
We present an eigenmode-free (EF) method-of-lines (MoL) formulation for the fast synthesis of reconfigurable graphene  metasurfaces (MS). As the complexity of MSs increases, analysis by full-wave methods becomes challenging. The MoL is a considerably faster semi-analytical method where the electromagnetic equations
are solved analytically along the direction perpendicular to the MS layers and numerically on the MS plane, thereby substantially decreasing the degrees of freedom (DoFs). In existing MoL formulations, the eigendecomposition of the system matrix is calculated numerically, which becomes computationally demanding for MSs with larger cross-sections. To overcome this limitation, we introduce an EF MoL that calculates the S-parameter matrix by analytical closed-form expressions. We demonstrate the potential of the EF MoL by analyzing a reconfigurable graphene MS absorber. The EF MoL shows excellent agreement in the absorbance and is two orders of magnitude faster than the finite element method.
\end{abstract}


\section{Introduction}
Metasurfaces (MS) can efficiently control  propagating waves and have gained increasing interest due to their low profile and planar form. Reconfigurable MSs are particularly important as they can switch between different functionalities~\cite{PRA}.  Various reconfigurable MSs have recently been introduced, with reconfigurability achieved using PIN diodes or tunable materials, such as graphene. Graphene's conductivity is dominated by the intraband term in the THz frequencies and may be tuned by varying the Fermi level, by means of electrical gating~\cite{PRA}. Although reconfigurable MSs can be accurately analyzed by full-wave methods, such as the finite element method (FEM), they become computationally expensive as the size and number of states increase.

The MoL presents an appealing alternative to full-wave methods for fast MS synthesis~\cite{Mehr}. It is a semi-analytical methodin which the wave equations are solved analytically along the direction perpendicular to the MS  and numerically on the MS plane, thus alleviating the need to discretize the entire 3D domain. The MoL has been used to analyze multilayer graphene structures~\cite{Mehr}, conformal antennas~\cite{Alu}, and mmWave MSs~\cite{Passia2025}. Existing MoL approaches require numerically solving an eigendecomposition problem, which becomes computationally demanding for devices with a large cross-section and/or very fine features, thus necessitating a finer discretization on the transverse plane. 

We present an EF MoL formulation that utilizes closed-form analytical expressions to form the eigendecomposition. Its benefits are twofold: (i) the complexity of calculating the transfer matrix is theoretically reduced, and (ii) the numerical accuracy of the MoL is improved.  We benchmark against the FEM for the analysis a reconfigurable graphene MS absorber.  The EF MoL shows excellent agreement in the absorbance and is two order of magnitude faster than the FEM.


\section{Eigenmode-free Method of Lines}

\subsection{MoL formulation}

The MoL formulation employs Maxwell's normalized equations expressed in terms of the tangential electric and magnetic field components. The coordinates $x,y,z$ are replaced by $x'{=}k_0x$, $y'{=}k_0y$, $z'{=}k_0z$, and the magnetic field component intensities $[\tilde{H}_x,\tilde{H}_y,\tilde{H}_z]$  by $\tilde{H}_x{=}Z_0 H_x$, $\tilde{H}_y{=}Z_0 H_y$, $\tilde{H}_z{=}Z_0 H_z$. We substitute the longitudinal components $E_z,\tilde{H}_z$ with the transverse ones~\cite{Passia2025}:
\begin{equation} \label{eq:1} \small
\begin{split}
\frac{\partial E_x}{\partial z'} &= -j  {\tilde{H}_y} -\frac{j}{ \varepsilon_r} \frac{\partial^2 \tilde{H}_y}{\partial {x'}^2} + \frac{j}{ \varepsilon_r} \frac{\partial^2 \tilde{H}_x}{\partial x' \partial y'},\\
\frac{\partial E_y}{\partial z'} &= +j {\tilde{H}_x} +\frac{j}{\varepsilon_r} \frac{\partial^2 \tilde{H}_x}{\partial {y'}^2} - \frac{j}{\varepsilon_r} \frac{\partial^2 \tilde{H}_y}{\partial y' \partial x'},\\
\frac{\partial \tilde{H}_x}{\partial z'} &= +j  \varepsilon_r {E_y} +{j} \frac{\partial^2 E_y}{\partial {x'}^2} - {j} \frac{\partial^2 E_x}{\partial x' \partial y'},\\
\frac{\partial \tilde{H}_y}{\partial z'} &= -j\varepsilon_r {E_x} -{j} \frac{\partial^2 E_x}{\partial {y'}^2} + {j} \frac{\partial^2 E_y}{\partial y' \partial x'}.\\
\end{split}    
\end{equation}
We consider a 2-D discretization scheme based on the Yee cell~\cite{Yee} on the $x{-}y$ plane. The grid size is $n_x {\times} n_y$, with $n_x$, $n_y$ the number of cells along the $x$ and $y$ axes.  
Periodic boundary conditions (PBCs) are imposed along the $x$ and $y$ axes. 
The discretized version of (\ref{eq:1}) is~\cite{Passia2025}:
\begin{equation}\label{eq:system_equation} \small
\frac{\partial}{\partial z} \begin{bmatrix}
    \mathbf{E_x}\\\mathbf{E_y}\\\mathbf{\tilde{H}_x}\\\mathbf{\tilde{H}_y}
\end{bmatrix} = j \begin{bmatrix}
    \mathbf{0} & \mathbf{Z}\\
    \mathbf{Y} & \mathbf{0}\\
\end{bmatrix} \begin{bmatrix}    \mathbf{E_x}\\\mathbf{E_y}\\\mathbf{\tilde{H}_x}\\\mathbf{\tilde{H}_y}
\end{bmatrix} = j \mathbf{A_s} \begin{bmatrix} \mathbf{E_x}\\\mathbf{E_y}\\\mathbf{\tilde{H}_x}\\\mathbf{\tilde{H}_y}
\end{bmatrix}.
\end{equation}
The solution to the above equation is a matrix exponential:
\begin{equation}\begin{bmatrix} \small
    \mathbf{E_x}(z)\\\mathbf{E_y}(z)\\\mathbf{\tilde{H}_x}(z)\\\mathbf{\tilde{H}_y}(z)
\end{bmatrix} = e^{\mathbf{A_s} k_0 z} \begin{bmatrix}
    \mathbf{E_{x,0}}\\\mathbf{E_{y,0}}\\\mathbf{\tilde{H}_{x,0}}\\\mathbf{\tilde{H}_{y,0}}
\end{bmatrix}.
\end{equation}
The term $e^{\mathbf{A_s}k_0 z}$ is the transfer matrix and $\mathbf{E_{x/y,0}}$ and $\mathbf{\tilde{H}_{x/y,0}}$ are the tangential field components on the $z$=0 plane.  To ensure numerical stability, an S-matrix approach is adopted instead of a T-matrix~\cite{Li}. The S-matrix of each dielectric layer is formed, suitable BCs are applied on the interfaces, and the S-parameter matrix of the entire device is obtained by connecting consecutive layers through the Redheffer star product~\cite{Rumpf2011}. The absorbance is calculated from the reflection and transmission coefficients of the entire MS as in~\cite{Passia2025}.

\subsection{Eigenmode-free approach}

We calculate the transfer matrix and consequently the S-parameter matrix of each layer using analytical closed-form expressions~\cite{TAP}.  Assuming a uniform rectangular grid for the disretization and a layer of spatially-invariant dielectric properties, the matrix $\mathbf{ZY}$ takes the following form:
\begin{equation} \label{eq:block-circulant-zy} \small
    \mathbf{ZY} {=} 
    \begin{bmatrix}
        \mathbf{P}_{xx}\mathbf{Q}_{xx} - \mathbf{M}_{xy}\mathbf{N}_{yx} & \mathbf{0}
        \\
        \mathbf{0} & \mathbf{P}_{yy}\mathbf{Q}_{yy} - \mathbf{M}_{yx}\mathbf{N}_{xy}
    \end{bmatrix}, 
\end{equation}
where the $\mathbf{P_{xx}, P_{yy}, M_{xy}, N_{yx}, P_{yy}, Q_{yy}, M_{yx}, N_{xy}}$ are submatrices of $\mathbf{Z}$ and $\mathbf{Y}$.
Since all submatrices are block-circulant with circulant blocks (BCCB), matrix $\mathbf{ZY}$ is diagonalizable by the 2D DFT~\cite{TAP}:
\begin{equation} \label{eq:fourer-diagon-zy} \small 
    \mathbf{ZY} {=} 
    \begin{bmatrix}
        \mathbf{U}_{n_x,n_y} & \mathbf{0}
        \\
        \mathbf{0} & \mathbf{U}_{n_x,n_y}
    \end{bmatrix}
    \!\!\!
    \begin{bmatrix}
        \mathbf{\Lambda}_{11} & \mathbf{0}
        \\
        \mathbf{0} & \mathbf{\Lambda}_{22}
    \end{bmatrix}
    \!\!\!
    \begin{bmatrix}
        \mathbf{U}_{n_x,n_y} & \mathbf{0}
        \\
        \mathbf{0} & \mathbf{U}_{n_x,n_y}
    \end{bmatrix}^{\top}\!\!,
\end{equation}
where $\mathbf{U}_{n_x,n_y} \triangleq \mathbf{F}_{n_x} \otimes \mathbf{F}_{n_y}$, $\mathbf{F}_n$ is the DFT matrix of size $n$ and $\otimes$ denotes the Kronecker product. The eigenvalues are computed as the 2D DFT of any column of matrix $\mathbf{P}_{xx}\mathbf{Q}_{xx} - \mathbf{M}_{xy}\mathbf{N}_{yx}$ for $\mathbf{\Lambda}_{11}$ and matrix $\mathbf{P}_{yy}\mathbf{Q}_{yy} - \mathbf{M}_{yx}\mathbf{N}_{xy}$ for $\mathbf{\Lambda}_{22}$, respectively.
This analytical solution  has complexity $\mathrm{O}(n_x n_y \log \left( n_x n_y\right))$ - with complexity $\mathrm{O}(n^2)$ for explicitly forming  eigenvector-related matrices, which is significantly lower than the $\mathrm{O}(n^3)$ needed for a dense eigensolver.

\subsection{Introducing graphene as a boundary condition}

Graphene induces a surface current density, $\mathbf{J_s} = \sigma_{g} \mathbf{E_{t,g}}$, where $\sigma_g$ is the graphene conductivity and $\mathbf{E_{t,g}}$ are the tangential electric field components.  The graphene's conductivity is expressed by Kubo's formula. The graphene conductivity is $
    \sigma_{g} (\omega){=}\frac{-j \mu_c e^2}{\pi \hbar (\omega-j \tau^{-1})}$,
where $\mu_c$ is the Fermi level, $\tau{=}\SI{3.28}{ps}$ is the relaxation time for the intraband transition, $e$ is the electron charge, and $\hbar$ is the reduced Planck’s constant. 
The graphene surface current density boundary condition (SCDBC) is introduced in the MoL, as any other SCDBC~\cite{Passia2025}.

\section{Results}

\begin{figure}
	\centering
\includegraphics[width=0.41\columnwidth,trim=0cm 0cm 0cm 0cm,clip]{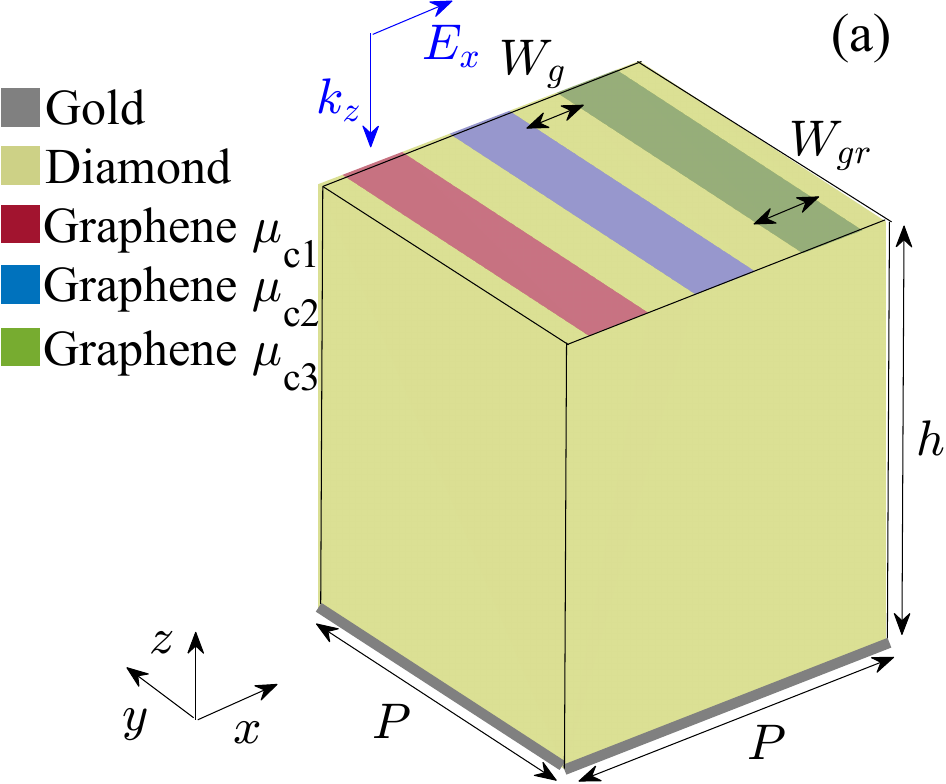}     
\includegraphics[width=0.57\columnwidth,trim=0cm 0cm 0cm 0cm,clip]{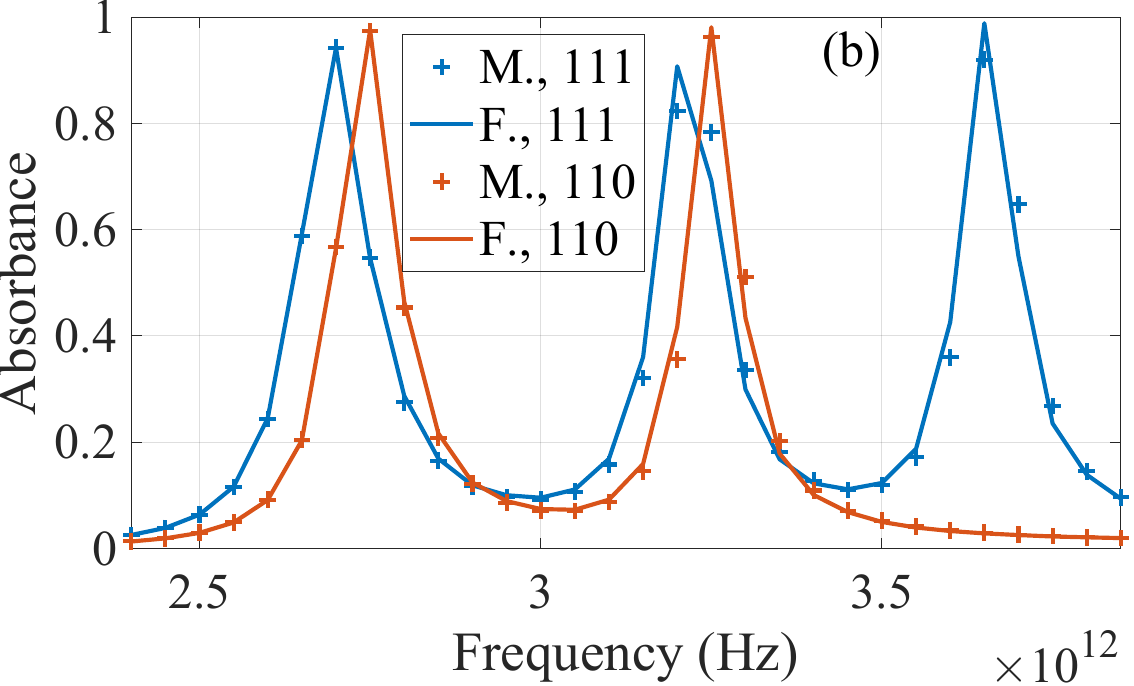}       	\caption{(a) Configuration and (b) Absorbance of the reconfigurable MS absorber, via the EF MoL (M.) and FEM (F.) for two states: 111 and 110.}
	  \label{fig:graph}
\end{figure}

We analyze the reconfigurable graphene MS absorber of~\cite{PRA} using the eigenmode-free MoL and compare the absorbance with that of the FEM (implemented in COMSOL Multiphysics). Three graphene strips of width $W_{gr}{=} \qty{1.1111}{\micro\metre}$, whose Fermi level can be tuned independently, are located on a gold-backed diamond substrate of thickness $h{=}\qty{7.2614}{\micro\metre}$ and refractive index $n{=}2.41$.  The unit cell size is $P{=}\qty{6}{\micro\metre}$ and the gap between the strips $W_g=\qty{0.8889}{\micro\metre}$ (Fig.~\ref{fig:graph}(a)). The absorber is excited by a normally incident plane wave, with the electric field oriented along the $x$ axis. By applying the following Fermi levels to each graphene strip: $\mu_{c,1}{=}\SI{0.088}{eV}$, $\mu_{c,2}{=}\SI{0.118}{eV}$, $\mu_{c,3}{=}\SI{0.148}{eV}$, three resonances appear in the absorbance response (Fig.~\ref{fig:graph}(b)), corresponding to state 111. If the Fermi level of the third patch is set to zero, two resonances appear, state 110 in Fig.~\ref{fig:graph}(b).

In the EF MoL, a cell grid of $n_x {\times} n_y = 27 {\times} 27$ is adopted. Graphene is introduced as a SCDBC in both methods. In the FEM gold is introduced as PEC, whereas in the MoL as a SCDBC, with $\sigma{=}1/R_s$ and $R_s{=}\SI{0.01}{$\Omega$/sq}$, which yields results identical to PEC. PBCs are applied to the side boundaries.  We compare the EF MoL response to the FEM in Fig.~\ref{fig:graph}(b), and excellent agreement is observed for both states. The MoL has $2n_x {\times} n_y{=}\SI{1 458}{DoFs}$ whereas the FEM has \SI{2 823 021}{DoFs}, assuming linear triangular and tetrahedral elements as fine as the MoL grid. The MoL computes the response of both states in approximately \SI{6}{s} per frequency, \SI[quantity-product={}]{380}{\times} faster than the \SI{38}{min} per frequency needed by the FEM.  Changing the Fermi level of the graphene strips requires fully re-simulating the FEM, whereas the eigendecomposition steps of the MoL are computed only once per frequency.


\section{Conclusion}
The introduced EF MoL accurately captures the graphene MS response and is \SI[quantity-product={}]{380}{\times} faster than the FEM, enabling the fast synthesis of larger reconfigurable MSs.

\end{document}